\documentclass[aps,prl,reprint,superscriptaddress,twocolumn]{revtex4-2}

\usepackage{amssymb}
\usepackage{amsmath}
\usepackage{graphicx}
\usepackage{epstopdf}
\usepackage[colorlinks, linkcolor=blue, urlcolor=blue, anchorcolor=blue, citecolor=blue]{hyperref}

\begin{document}
\title{Experimental Realization of the Rabi-Hubbard Model with Trapped Ions}

\author{Q.-X. Mei}
\thanks{These authors contribute equally to this work}%
\affiliation{Center for Quantum Information, Institute for Interdisciplinary Information Sciences, Tsinghua University, Beijing 100084, P. R. China}
\author{B.-W. Li}
\thanks{These authors contribute equally to this work}%
\affiliation{Center for Quantum Information, Institute for Interdisciplinary Information Sciences, Tsinghua University, Beijing 100084, P. R. China}
\author{Y.-K. Wu}
\thanks{These authors contribute equally to this work}%
\affiliation{Center for Quantum Information, Institute for Interdisciplinary Information Sciences, Tsinghua University, Beijing 100084, P. R. China}
\author{M.-L. Cai}
\affiliation{Center for Quantum Information, Institute for Interdisciplinary Information Sciences, Tsinghua University, Beijing 100084, P. R. China}
\affiliation{HYQ Co., Ltd., Beijing, 100176, P. R. China}
\author{Y. Wang}
\affiliation{Center for Quantum Information, Institute for Interdisciplinary Information Sciences, Tsinghua University, Beijing 100084, P. R. China}
\author{L. Yao}
\affiliation{Center for Quantum Information, Institute for Interdisciplinary Information Sciences, Tsinghua University, Beijing 100084, P. R. China}
\affiliation{HYQ Co., Ltd., Beijing, 100176, P. R. China}
\author{Z.-C. Zhou}
\affiliation{Center for Quantum Information, Institute for Interdisciplinary Information Sciences, Tsinghua University, Beijing 100084, P. R. China}
\author{L.-M. Duan}
\email{lmduan@tsinghua.edu.cn}
\affiliation{Center for Quantum Information, Institute for Interdisciplinary Information Sciences, Tsinghua University, Beijing 100084, P. R. China}

\begin{abstract}
Quantum simulation provides important tools in studying strongly correlated many-body systems with controllable parameters. As a hybrid of two fundamental models in quantum optics and in condensed matter physics, the Rabi-Hubbard model demonstrates rich physics through the competition between local spin-boson interactions and long-range boson hopping. Here we report an experimental realization of the Rabi-Hubbard model using up to $16$ trapped ions and present a controlled study of its equilibrium properties and quantum dynamics. We observe the ground-state quantum phase transition by slowly quenching the coupling strength, and measure the quantum dynamical evolution in various parameter regimes. With the magnetization and the spin-spin correlation as probes, we verify the prediction of the model Hamiltonian by comparing theoretical results in small system sizes with experimental observations. For larger-size systems of $16$ ions and $16$ phonon modes, the effective Hilbert space dimension exceeds $2^{57}$, whose dynamics is intractable for classical supercomputers.
\end{abstract}

\maketitle

Hubbard model is a fundamental model in many-body physics, with rich physical phenomena arising from two competing effects: the on-site repulsion and the hopping between different sites \cite{tasaki2020introduction}. One of its natural generalizations to the spin-boson coupled system is the Rabi-Hubbard (RH) model \cite{schiro2012phase,hwang2013largescale,zhu2013dispersive} where the on-site interaction is replaced by a quantum Rabi model Hamiltonian \cite{forn-diaz2019ultrastrong}, a fundamental model in quantum optics describing interaction of a spin with a bosonic mode. The RH model breaks the $U(1)$ symmetry (particle number conservation) that appears in the other generalizations like Bose-Hubbard model \cite{fisher1989boson,jaksch1998cold,bloch2008manybody} and Jaynes-Cummings-Hubbard (JCH) model \cite{greentree2006quantum,angelakis2007photonblockadeinduced}, and thus shows nontrivial distinctions in its ground state or general dynamics. Although important analytical and numerical progress has been achieved in understanding its properties \cite{zheng2011importance,schiro2012phase,hwang2013largescale,zhu2013dispersive,flottat2016quantum,schiro2016exotic}, the RH model has not yet been realized in the cavity QED system where it was first proposed due to the experimental difficulty. This motivates one to realize and experimentally probe the RH model using other controllable physical systems through the idea of quantum simulation \cite{georgescu2014quantum,cirac2012goals}.

As the scale and the controllability of quantum devices develop, quantum simulation is becoming increasingly important in studying strongly correlated many-body systems \cite{georgescu2014quantum,cirac2012goals}. As one of the leading platforms for quantum simulation, trapped ions possess long coherence time, convenient initialization and readout \cite{leibfried2003quantum}. Furthermore, the trapped ion system is intrinsically equipped with laser-coupled spin and bosonic degrees of freedom \cite{leibfried2003quantum}, which makes it an excellent candidate to simulate light-matter interaction Hamiltonian. Previously, quantum simulation of many-body spin models \cite{monroe2021programmable}, Dicke model \cite{safavi-naini2018verification}, quantum Rabi model of a single ion \cite{lv2018quantum,cai2021observation}, JCH model for two \cite{toyoda2013experimental,ohira2021Blockade} and three \cite{debnath2018observation} ions have been demonstrated in this system. Here, we perform quantum simulation of the RH model for the first time with up to 16 ions and explore its equilibrium phase transition \cite{sachdev2000quantum} and quantum dynamical properties \cite{polkovnikov2011colloquium} using spin observables. Compared with the spin models where phonons are only virtually excited \cite{monroe2021programmable}, our inclusion of phonon modes in realization of the RH model greatly enlarges the effective dimension of the Hilbert space and thus demonstrates quantum simulation results that are intractable for the available classical computers.

\begin{figure}[htbp]
	\centering
	\includegraphics[width=\linewidth]{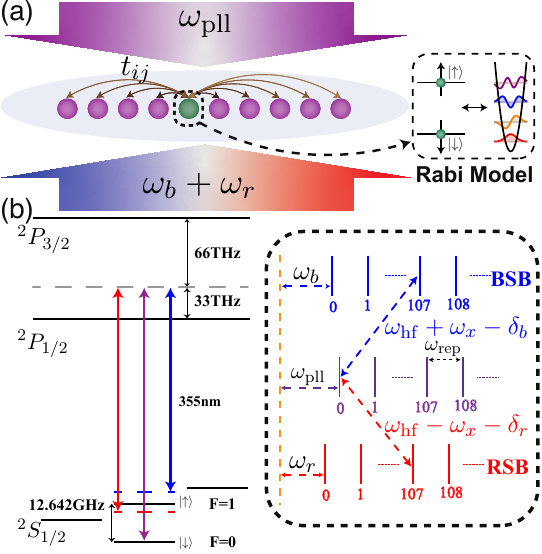}
	\caption { Schematic of the experiment. (a) We use two global Raman laser beams to create a Rabi-Hubbard model Hamiltonian on an ion chain. Two frequency components are used to drive the blue and the red phonon sidebands (BSB and RSB) simultaneously. The frequency of the Raman pairs is locked by a phase-locked loop (PLL). The hopping rates among different sites $t_{ij}$ are determined by the inter-ion spacings, and the local quantum Rabi model Hamiltonian is controlled by the amplitudes and the frequencies of the driving lasers. (b) We use $355\,$nm pulsed laser beams with a frequency comb structure to bridge the Raman transitions of the qubits \cite{islam2014beat}. The large energy splitting of $\omega_{\mathrm{hf}}=2\pi\times 12.6\,$GHz is covered by about $107$ teeth of the frequency comb so we only need frequency shifts on the order of tens of MHz to set suitable detuning for the bichromatic Raman beams.}
\end{figure}

\emph{Long-range Rabi-Hubbard model.} We use a chain of trapped ${}^{171}\mathrm{Yb}^+$ ions to simulate the RH model. Our experimental setup is shown schematically in Fig.~1. The on-site quantum Rabi model Hamiltonian is generated through global bichromatic Raman laser beams \cite{cai2021observation,lv2018quantum} which couple the internal qubit states $|\downarrow\rangle\equiv |S_{1/2},F=0,m_F=0\rangle$, $|\uparrow\rangle\equiv |S_{1/2},F=1,m_F=0\rangle$ with the local transverse oscillation of the ions. Furthermore, the Coulomb interaction between the ions couples these local oscillation modes together and finally gives us an RH Hamiltonian
\begin{align}
H =& \sum_i \left[\frac{\omega_0}{2} \sigma_z^i + \omega_i a_i^\dag a_i + g \sigma_x^i (a_i + a_i^\dag)\right] \nonumber\\
&\qquad + \sum_{i<j} t_{ij} (a_i^\dag a_j + a_j^\dag a_i),
\end{align}
where $\sigma_x^i$ and $\sigma_z^i$ are Pauli operators for the spin $i$ and $a_i$ and $a_i^\dag$ the annihilation and creation operators of the corresponding local phonon mode. The spin frequency $\omega_0$ is set by the detuning of the global bichromatic laser beams, the spin-phonon coupling $g$ by the amplitudes, the phonon hopping term $t_{ij}$ by the ion spacings, and the local phonon frequency $\omega_i$ by both the laser detuning and the Coulomb interaction and thus becomes inhomogeneous (see Supplementary Materials for details \cite{supplementary}). Compared with the original RH model \cite{schiro2012phase}, our Hamiltonian has long-range hopping decaying inverse cubically with the distance.

\begin{figure*}[htbp]
	\centering
	\includegraphics[width=\linewidth]{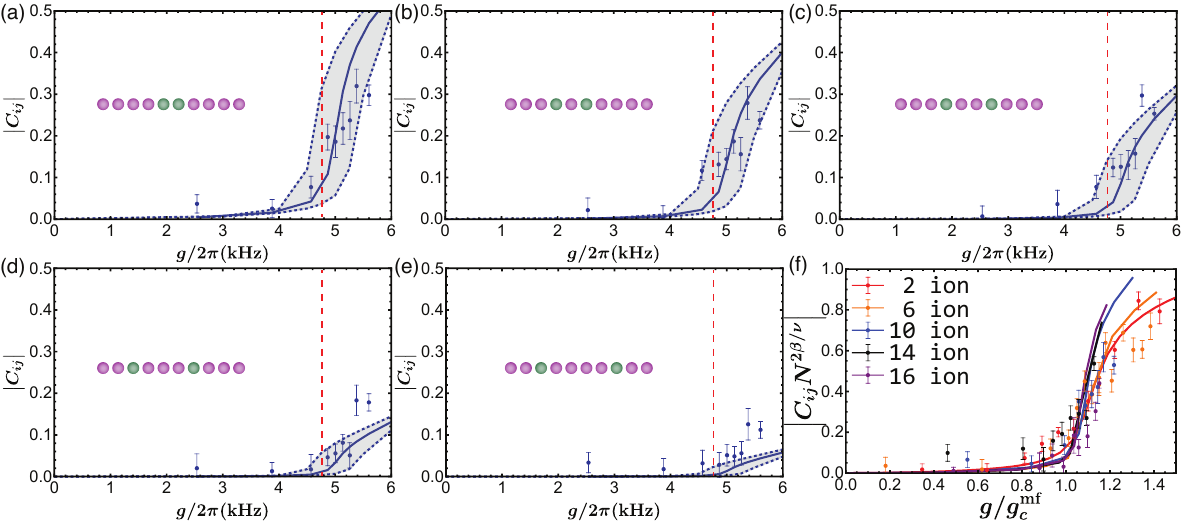}
	\caption {Quantum phase transition under slow quench. We start from the ground state $|\Psi_0\rangle\equiv |\downarrow,0\rangle^{\otimes N}$ with the on-site spin-phonon coupling $g=0$, and then slowly tune up the coupling across the predicted critical point. (a)-(e) Spin-spin correlations $C_{ij}\equiv \langle\sigma_x^i\sigma_x^j\rangle - \langle\sigma_x^i\rangle \langle\sigma_x^j\rangle$ for various ion pairs in an $N=10$ chain versus the coupling $g$ after slow quench. For small $g$, the correlation remains close to zero apart from small detection errors; once $g$ is tuned across a critical point (indicated by the vertical dashed line as the numerically computed value $g_c\approx 1.03 g_c^{\mathrm{mf}}$), the spin-spin correlation increases rapidly, which indicates a quantum phase transition. The solid line is the theoretical ground-state value from the DMRG calculation, and the shaded region between the dashed lines represent the theoretical results under a shift of $\pm 300\,$Hz in the trap frequency. (f) The nearest-neighbor spin-spin correlation for two central ions in a chain of $2-16$ ions (dots with error bars representing one standard deviation) and the corresponding theoretical ground-state values from the DMRG calculation (solid lines). Here we normalize the horizontal axis by the mean-field transition point $g_c^{\mathrm{mf}}$, and scale the vertical axis by $N^{2\beta/\nu}$ where $\beta=1/8$ and $\nu=1$ are two critical exponents. Theoretically, the rise of the curves becomes sharper near the predicted transition point as $N$ increases. Although this is less clear from the experimental data due to the noise and errors including the violation of adiabaticity and decoherence, the overall tendency between the theoretical and the experimental results still agree with each other for different system sizes (see Supplementary Materials for individual plots of each ion number $N$ \cite{supplementary}).}
\end{figure*}

\emph{Equilibrium quantum phase transition.} First we study the quantum phase transition in this model by slowly tuning the spin-phonon coupling across the critical point, as shown in Fig.~2. The RH model has two distinct phases in its ground state \cite{schiro2012phase}: at low phonon hopping rate and low spin-phonon coupling, the spin-spin correlation on distant sites vanishes, which is known as an incoherent phase; as the hopping and the coupling rates increase, long-range spin-spin correlation appears as the $Z_2$ symmetry spontaneously breaks, and the system enters a coherent phase. Here we start from zero spin-phonon coupling, for which the ground state can be easily prepared by sideband cooling of phonon modes into $|0\rangle$ and optical pumping of spins into $|\downarrow\rangle$. As the spin-phonon coupling $g$ increases, the ground state phase transition can be understood qualitatively by a mean-field analysis (see Supplementary Materials for details \cite{supplementary}): Diagonalizing the local phonon modes $\{a_i\}$ into collective modes $\{b_k\}$ and ignoring the quantum correlation between spin and phonon states, the only consistent solution at low $g$ is $\langle\sigma_x^i\rangle=0$ and $\langle b_k\rangle=0$; As $g$ goes up across a critical point $g_c^{\mathrm{mf}}=\sqrt{\omega_0 \delta_0}/2$ where $\delta_0$ is the lowest frequency of the collective modes, $\langle b_0\rangle$ can acquire a nonzero value, which in turn leads to nonzero $\langle \sigma_x^i\rangle$ for each spin and nonzero $\langle b_k\rangle$ for other modes.
In the experiment, we slowly tune up the coupling $g$ following an exponential function $g(t)=(1-e^{-t/\tau})g_{\mathrm{max}}$ where $\tau=1\,$ms is the largest quench time and $(1-1/e)g_{\mathrm{max}}$ above the critical point is the largest coupling rate. We expect the system to stay in the ground state adiabatically until close to the transition point where the energy gap closes in the thermodynamic limit. Nevertheless, this still allows us to observe the transition signal in the spin-spin correlation.

In our experiment, $g$ is limited by the available laser power, so we set small $\delta_0\approx 2\pi\times 2\,$kHz for an achievable critical point. In Fig.~2(a)-(e), we present the spin-spin correlation $C_{ij}\equiv \langle\sigma_x^i\sigma_x^j\rangle - \langle\sigma_x^i\rangle \langle\sigma_x^j\rangle$ for ion pairs with various distances in an $N=10$ chain. At low coupling, ideally the spin-spin correlation is vanishingly small. In the experiment, we measure the spin-spin correlation by rotating $\sigma_\phi=\sigma_x\cos\phi+\sigma_y\sin\phi$ into the $\sigma_z$ basis, scan the correlation with respect to $\phi$, and then extract the oscillation amplitude of this curve as $C_{ij}$ (see Supplementary Materials \cite{supplementary}). While this process removes the sensitivity to the relative phase between the lab frame and the interaction picture of the RH Hamiltonian, it causes a systematic error when $C_{ij}$ is close to zero since the fitted oscillation amplitude is always non-negative: The statistical fluctuation or drift in the device parameters will now result in a measured positive correlation in the low-$g$ regime, which is what we observe in these plots. As $g$ rises near the mean-field transition point $g_c^{\mathrm{mf}} = 2\pi\times 4.6\,$kHz, we observe quick increase in the spin-spin correlations which is a signature of the quantum phase transition. The experimental data agree well with the theoretical results calculated by the density-matrix-renormalization-group (DMRG) method \cite{schollwock2011densitymatrix} (solid lines), and the error comes from slow drifts in the trap frequencies (dashed lines for $\pm 300\,$Hz drifts), violation of the adiabatic condition, motional decoherence as well as SPAM errors (see Supplementary Materials for details \cite{supplementary}). Furthermore, we see that the correlation in the coherent phase decays slowly with the distance and persists over half a chain, which is characteristic for this phase. In Fig.~2(f) we plot the experimental and the theoretical results for ion numbers ranging from $2$ to $16$. Although the current experimental accuracy is not enough for a finite size scaling analysis, we see that the experimental results are consistent with a transition point $g_c\approx 1.03 g_c^{\mathrm{mf}}$ predicted by the DMRG calculation with critical exponents $\beta=1/8$ and $\nu=1$.
More details about this phase transition, choice of parameters and adiabaticity can be found in Supplementary Materials \cite{supplementary}. Also note that in Fig.~2(f), the range of data points for large $N$ is narrower than those for small $N$. This is because for the experimental data we choose, $g_c^{\mathrm{mf}}$ is about $40\%$ higher for the $N=16$ case than for the $N=2$ case so that the rescaled coupling $g/g_c^{\mathrm{mf}}$ decreases for the same spin-phonon coupling $g$. Besides, in the experiment we observe that the lifetime of the ion crystal under strong laser driving decreases with the increasing ion number, therefore for higher $N$ we need to use smaller $g$.

\begin{figure*}[htbp]
	\centering
	\includegraphics[width=\linewidth]{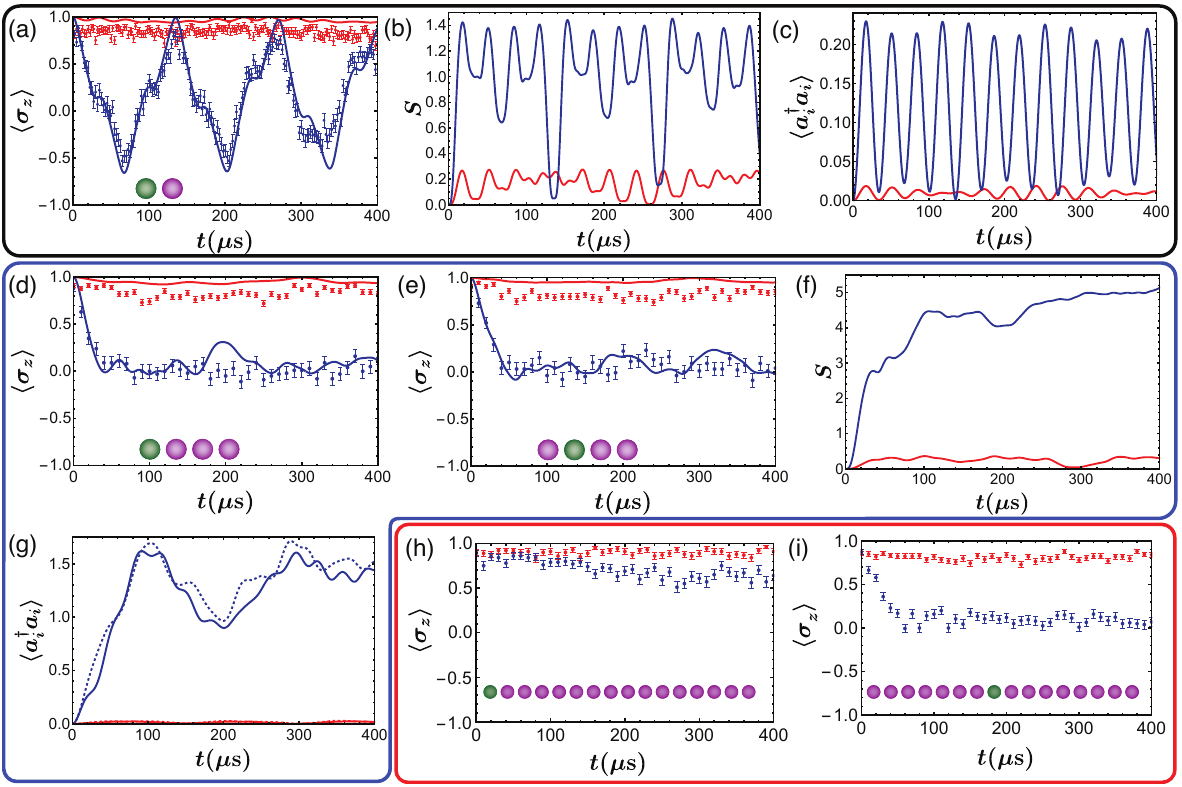}
	\caption {Generic spin dynamics. We initialize the system in $|\uparrow,0\rangle^{\otimes N}$, immediately turn on the RH Hamiltonian to evolve the system and measure $\langle\sigma_z^i(t)\rangle$ of individual spins. (a) Measured data (dots with error bars representing one standard deviation) and theoretical results from numerically integrating the Schrodinger equation (solid lines) for $N=2$ at $g=2\pi\times 2\,$kHz (red) and $2\pi\times 7\,$kHz (blue). The two ions are symmetric and hence only one is plotted. (d),(e) Similar results for $N=4$ at $g=2\pi\times 1\,$kHz (red) and $2\pi\times 6\,$kHz (blue) for an ion on the edge and in the center, respectively. (b),(f) The corresponding evolution of entanglement entropy $S(t)$ for $N=2$ and $N=4$ between the left and the right halves of the chain (with both spin and phonon states included) under the two coupling strengths. (c),(g) Corresponding theoretical results for local phonon numbers $\langle a_i^\dag(t) a_i(t)\rangle$. For $N=4$ the solid and the dashed lines are for the central and the edge sites, respectively. (h),(i) The measured dynamics for the edge and the central spins of an $N=16$ chain at $g=2\pi\times 1\,$kHz (red) and $2\pi\times 6\,$kHz (blue).}
\end{figure*}

\emph{Quantum dynamics.} Next we consider nonequilibrium quantum dynamics of the RH Hamiltonian. As mentioned above, the ground-state properties of tens of ions can be computed using the DMRG method because of the low amount of entanglement in the one-dimensional system \cite{schollwock2011densitymatrix}. However, such methods will no longer be applicable for quantum dynamics far from equilibrium. In Fig.~3 we initialize the system in $|\uparrow,0\rangle^{\otimes N}$ through sideband cooling and optical pumping followed by a global Raman $\pi$ pulse. Then we turn on the RH Hamiltonian and measure the evolution of $\langle \sigma_z^i(t)\rangle$ for individual ions. In Fig.~3(a), (d) and (e), we see the measured dynamics agree well with the theoretical results from direct numerical integration of the Schrodinger equation for small system sizes: At small $g$, the spins are barely affected by the phonon coupling and hopping, and thus stays near $\langle \sigma_z^i(t)\rangle\approx 1$ (the lower experimental curves mainly come from SPAM errors); for larger $g$, the phonon modes become more important and we observe oscillatory or damping behavior in the spin dynamics. In Supplementary Materials, we further show that this difference can be understood qualitatively from the stability of the system under the Holstein–Primakoff approximation \cite{supplementary}. In Fig.~3(b) and (f) we present the corresponding theoretical entanglement entropy between the left and the right halves of the chain (with both spin and phonon states included) and in Fig.~3(c) and g we plot the theoretical phonon numbers. All these theoretical results demonstrate the explicit involvement of the phonon modes in the dynamics we are studying. (Note that an entanglement entropy higher than $N/2$ proves that here the phonon states directly contribute to the entanglement and thus the dynamics are strikingly different from the Ising models simulated in earlier works \cite{georgescu2014quantum} where the phonon states are adiabatically eliminated.) For an average phonon number around $1.5$ as shown in Fig.~3(g), we need a cutoff of at least $6$ for the local phonon number to capture the dynamics, thus the dimension of the system scales as $[2\times(6+1)]^N$ (see Supplementary Materials for details as well as for possible simplification using collective modes rather than local phonon modes \cite{supplementary}). On the other hand, the quick increase in the entanglement entropy clearly shows that here the matrix-product-state-based methods will not be applicable \cite{schollwock2011densitymatrix}. Therefore, the evolution of the RH model under the system size and the coupling strength achieved in this work, such as the spin dynamics in Fig.~3(h) and (i) for $N=16$ ions, will generally be difficult to simulate by classical supercomputers \cite{arute2019quantum}: The dimension of $14^{16} \approx 2^{61}$ or, with the possible simplification in Supplementary Materials using collective modes \cite{supplementary}, of about $2^{57}$, corresponds to 57 spins and even writing down such a pure state would take thousands of PB memories; besides, the phonon frequencies on the order of $2\pi\times 50\,$kHz and the evolution time up to $400\,\mu$s require hundreds of layers of single-site and two-site unitary gates.

\emph{Conclusion.} In summary, we have reported the first experimental realization of the Rabi-Hubbard model and performed quantum simulation of both ground-state and dynamical properties of this model using a chain of up to 16 ions. We verify the simulation of the Hamiltonian by showing agreements between theories and experiments for quantum phase transition and for generic spin dynamics in small scales. We then perform quantum simulation in large scales that is generally intractable for classical supercomputers.
This experiment allows exploring rich ground-state and quantum dynamical properties of the RH model in future works, and showcases that the trapped ion system provides an ideal platform to probe and quantum simulate various spin-boson many-body models, which naturally arise and play
important roles in a number of physics fields \cite{schiro2012phase,hwang2013largescale,zhu2013dispersive,greentree2006quantum,angelakis2007photonblockadeinduced,hartmann2006strongly,kurcz2014hybrid}.

\begin{acknowledgments}
We thank X. Zhang and L. He for discussions. This work was supported by Tsinghua University Initiative Scientific Research Program, Beijing Academy of Quantum Information Sciences, National Key Research and Development Program of China, and Frontier Science Center for Quantum Information of the Ministry of Education of China. Y.-K. W. acknowledges support from Shuimu Tsinghua Scholar Program, International Postdoctoral Exchange Fellowship Program (Talent-Introduction Program) and the start-up fund from Tsinghua University.
\end{acknowledgments}

%\bibliographystyle{apsrev4-1-title}
%\bibliography{reference}

\end{document}

\title{Supplementary Materials for ``Experimental Realization of Rabi-Hubbard Model with Trapped Ions''}

\author{Q.-X. Mei}
\thanks{These authors contribute equally to this work}%
\affiliation{Center for Quantum Information, Institute for Interdisciplinary Information Sciences, Tsinghua University, Beijing 100084, P. R. China}
\author{B.-W. Li}
\thanks{These authors contribute equally to this work}%
\affiliation{Center for Quantum Information, Institute for Interdisciplinary Information Sciences, Tsinghua University, Beijing 100084, P. R. China}
\author{Y.-K. Wu}
\thanks{These authors contribute equally to this work}%
\affiliation{Center for Quantum Information, Institute for Interdisciplinary Information Sciences, Tsinghua University, Beijing 100084, P. R. China}
\author{M.-L. Cai}
\affiliation{Center for Quantum Information, Institute for Interdisciplinary Information Sciences, Tsinghua University, Beijing 100084, P. R. China}
\affiliation{HYQ Co., Ltd., Beijing, 100176, P. R. China}
\author{Y. Wang}
\affiliation{Center for Quantum Information, Institute for Interdisciplinary Information Sciences, Tsinghua University, Beijing 100084, P. R. China}
\author{L. Yao}
\affiliation{Center for Quantum Information, Institute for Interdisciplinary Information Sciences, Tsinghua University, Beijing 100084, P. R. China}
\affiliation{HYQ Co., Ltd., Beijing, 100176, P. R. China}
\author{Z.-C. Zhou}
\affiliation{Center for Quantum Information, Institute for Interdisciplinary Information Sciences, Tsinghua University, Beijing 100084, P. R. China}
\author{L.-M. Duan}
\email{lmduan@tsinghua.edu.cn}
\affiliation{Center for Quantum Information, Institute for Interdisciplinary Information Sciences, Tsinghua University, Beijing 100084, P. R. China}

\maketitle

\section{Experimental setup}
We use a linear Paul trap with segmented blade electrodes to confine a chain of $^{171}\mathrm{Yb}^+$ ions. The axial potential can be manipulated by 5 pairs of DC electrodes to get nearly uniform ion spacing in the middle. We set the average ion spacing to be about $5.4\,\mu$m, which is further calibrated using the methods in the following sections. The qubits are encoded in the $|\downarrow\rangle\equiv |S_{1/2},F=0,m_F=0\rangle$ and $|\uparrow\rangle\equiv |S_{1/2},F=1,m_F=0\rangle$ levels. The $|\downarrow\rangle$ state can be initialized by optical pumping using resonant $369.5\,$nm laser with $2.105\,$GHz EOM sideband, and the $|\uparrow\rangle$ state can be prepared by a further Raman $\pi$ pulse.

Two counter-propagating $355\,$nm Raman laser beams \cite{cai2021observation} at an angle of $45^\circ$ to both the $x$ and $y$ axes are focused into the shape of an ellipse with Gaussian waist sizes (lengths of the major axis and the minor axis where the intensity drops to $1/e^2$) around $300\,\mu\mathrm{m} \times 20\,\mu\mathrm{m}$ at the position of the ion chain. The major axis is aligned with the chain to achieve nearly uniform coupling for all the ions. This pulsed laser has a specially designed repetition rate of $\omega_{\mathrm{rep}}=2\pi\times 118.414\,$MHz such that the hyperfine splitting satisfies $\omega_{01}\approx (k+1/2)\omega_{\mathrm{rep}}/2$ where $k$ is an integer. Under this condition, the fourth order AC Stark shifts \cite{lee2016engineering} from the bichromatic Raman beams will largely cancel each other.

\begin{figure}
    \centering
    \includegraphics[width=0.6\textwidth]{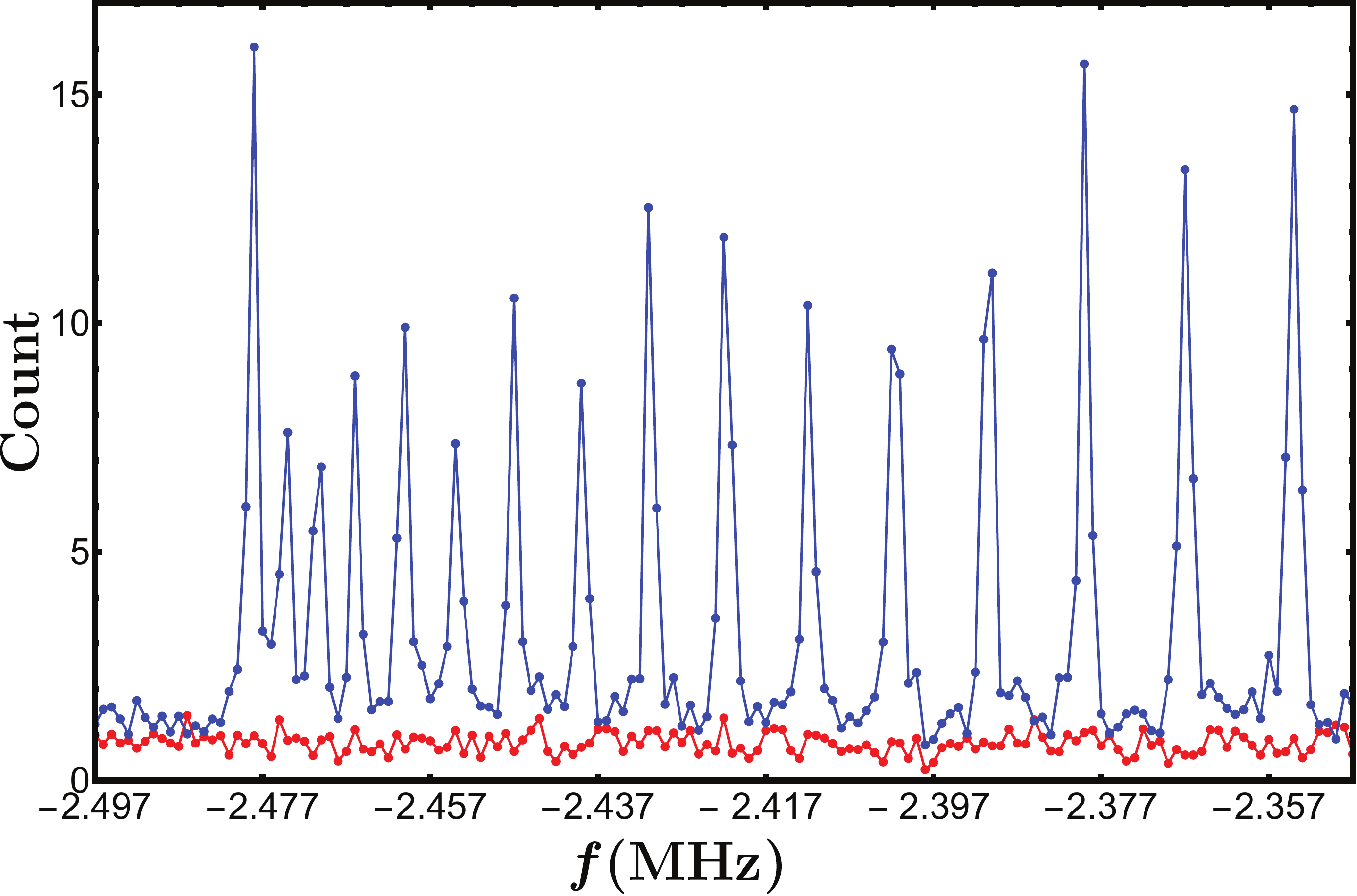}
    \caption{\label{fig:Cooling}Red sideband spectroscopy for $x$ modes of 16 ions before (blue) and after (red) sideband cooling.}
\end{figure}

The motional ground state is prepared first by $1\,$ms Doppler cooling using $369.5\,$nm laser, followed by $2\,$ms Raman sideband cooling \cite{leibfried2003quantum}. The $x$ and $y$ modes are separated by about $300\,$kHz, each spans a range of about $120\,$kHz for $N=16$ ions under the chosen ion spacings. Although we only use the $x$ modes in the later experiment, here we also cool the $y$ modes to suppress off-resonant coupling. We select 4 modes for the $x$ and $y$ directions each in the sideband cooling sequence and the final phonon numbers in all these modes are below $\bar{n}=0.1$, as shown in Fig.~\ref{fig:Cooling}.

To estimate the coherence time, we measure the Ramsey fringes by applying carrier or blue-sideband $\pi/2$ pulses using the Raman laser beams on a single trapped ion. This gives us the spin coherence time of about $20\,$ms and the motional coherence time of about $3\,$ms.
\section{Simulating Rabi-Hubbard Hamiltonian}
The Rabi-Hubbard model is simulated by the bichromatic Raman laser beams together with the Coulomb interaction between the ions. For small oscillation in the transverse $x$ direction, we have
\begin{equation}
H_m = \frac{1}{2} \sum_i \left(m\omega_x^2 - \frac{e^2}{4\pi\epsilon_0} \sum_{j\ne i} \frac{1}{z_{ij}^3} \right) x_i^2 + \sum_{i<j} \frac{e^2}{4\pi\epsilon_0} \frac{1}{z_{ij}^3} x_i x_j + \sum_i \frac{p_i^2}{2m},
\end{equation}
where $z_{ij}=|z_i-z_j|$ is the distance between the equilibrium positions of two ions.
If we define local trap frequency $\omega_i = \sqrt{\omega_x^2 - (e^2/4\pi\epsilon_0 m)\sum_{j\ne i}1/z_{ij}^3}$ and quantize the local oscillation correspondingly as $x_i=\sqrt{\hbar/2m\omega_i}(a_i+a_i^\dag)$ and $p_i=i\sqrt{\hbar m \omega_i / 2} (a_i^\dag - a_i)$, we obtain
\begin{equation}
H_m = \sum_i \omega_i a_i^\dag a_i + \sum_{i<j} t_{ij} (a_i + a_i^\dag)(a_j + a_j^\dag), \label{eq:motional}
\end{equation}
where $t_{ij}=e^2/(8\pi\epsilon_0 m \sqrt{\omega_i\omega_j} z_{ij}^3)$ and we have set $\hbar=1$ for convenience. In the case $e^2/4\pi\epsilon_0 m \omega_x^2 z_{ij}^3 \ll 1$, we get $t_{ij}\approx e^2/8\pi\epsilon_0 m \omega_x z_{ij}^3$ and $\omega_i \approx \omega_x - (e^2/8\pi\epsilon_0 m \omega_x) \sum_{j\ne i} 1 / z_{ij}^3$, and we can use the rotating-wave approximation to throw away the counter-rotating terms in Eq.~(\ref{eq:motional}). For a more accurate derivation, we note that in a frame rotating at the frequency $\omega_x$ for each local mode, the counter-rotating terms can be represented by an effective Hamiltonian \cite{goldman2014periodically} $H'=-(1/2\omega_x)\sum_i(\sum_{j\ne i}t_{ij}^2)a_i^\dag a_i - (1/2\omega_x)\sum_{i<j} (\sum_{k\ne i,j} t_{ik} t_{jk})(a_i^\dag a_j + a_j^\dag a_i)$. Therefore, after rotating wave approximation we have
\begin{equation}
H_m = \sum_i \tilde{\omega}_i a_i^\dag a_i + \sum_{i<j} \tilde{t}_{ij} (a_i^\dag a_j + a_j^\dag a_i),  \label{eq:motional_RWA}
\end{equation}
with
\begin{equation}
\tilde{\omega}_i = \omega_i - \frac{1}{2\omega_x}\sum_{j\ne i}t_{ij}^2,
\end{equation}
\begin{equation}
\tilde{t}_{ij} = t_{ij} - \frac{1}{2\omega_x} \sum_{k\ne i,j} t_{ik} t_{jk},
\end{equation}
and
\begin{equation}
\omega_i = \sqrt{\omega_x^2 - \frac{e^2}{4\pi\epsilon_0 m}\sum_{j\ne i}\frac{1}{z_{ij}^3}}, \label{eq:local_freq_low_order}
\end{equation}
\begin{equation}
t_{ij}=\frac{e^2}{8\pi\epsilon_0 m \sqrt{\omega_i\omega_j} z_{ij}^3}. \label{eq:hopping_low_order}
\end{equation}

Now we add the spin Hamiltonian $H_s=\sum_i(\omega_{\mathrm{hf}}/2)\sigma_z^i$ and the coupling Hamiltonian $H_{b(r)}=\sum_i\Omega_{b(r)}\cos[k_{b(r)}x_i-\omega_{b(r)}t+\phi_{b(r)}]\sigma_x^i$, and choose $\Omega_b=\Omega_r=\Omega$ and $\omega_b=\omega_{\mathrm{hf}}+\omega_x-\delta_b$, $\omega_r=\omega_{\mathrm{hf}}-\omega_x-\delta_r$ \cite{cai2021observation}. Then in an interaction picture with $H_0=\sum_i (\omega_{\mathrm{hf}}/2)\sigma_z^i + \sum_i \omega_x a_i^\dag a_i$, we get
\begin{equation}
H_I = \sum_i (\tilde{\omega}_i - \omega_x) a_i^\dag a_i + \sum_{i<j} \tilde{t}_{ij} (a_i^\dag a_j + a_j^\dag a_i) + \sum_i \left[ \frac{\eta_i\Omega}{2} \sigma_+^i (a_i e^{i\delta_r t} + a_i^\dag e^{i\delta_b t}) + h.c.\right],
\end{equation}
where $\eta_i\equiv k_{b(r)}\sqrt{\hbar/2 m \omega_i}$ are Lamb-Dicke parameters and are nearly uniform for all the ions.
If we further go into an interaction picture with $H_0' = -(\delta_b+\delta_r)/4 \sum_i\sigma_z^i - (\delta_b-\delta_r)/2\sum_i a_i^\dag a_i$, we finally get
\begin{equation}
H_I' =  \sum_i \left[\frac{\delta_b+\delta_r}{4}\sigma_z^i + \left(\tilde{\omega}_i - \omega_x + \frac{\delta_b - \delta_r}{2}\right) a_i^\dag a_i + \frac{\eta_i\Omega}{2} \sigma_x^i (a_i + a_i^\dag)\right]  + \sum_{i<j} \tilde{t}_{ij} (a_i^\dag a_j + a_j^\dag a_i), \label{eq:H_final}
\end{equation}
which is the desired Rabi-Hubbard model. In the main text, we simplify the notation by denoting $(\delta_b+\delta_r)/2$ as $\omega_0$, $\tilde{\omega}_i-\omega_x+(\delta_b-\delta_r)/2$ as $\omega_i$, $\eta_i\Omega/2$ as $g$ and $\tilde{t}_{ij}$ as $t_{ij}$.

\section{Calibrating experimental parameters}
\subsection{Ion distance}
\begin{figure}
    \centering
    \includegraphics[width=0.8\textwidth]{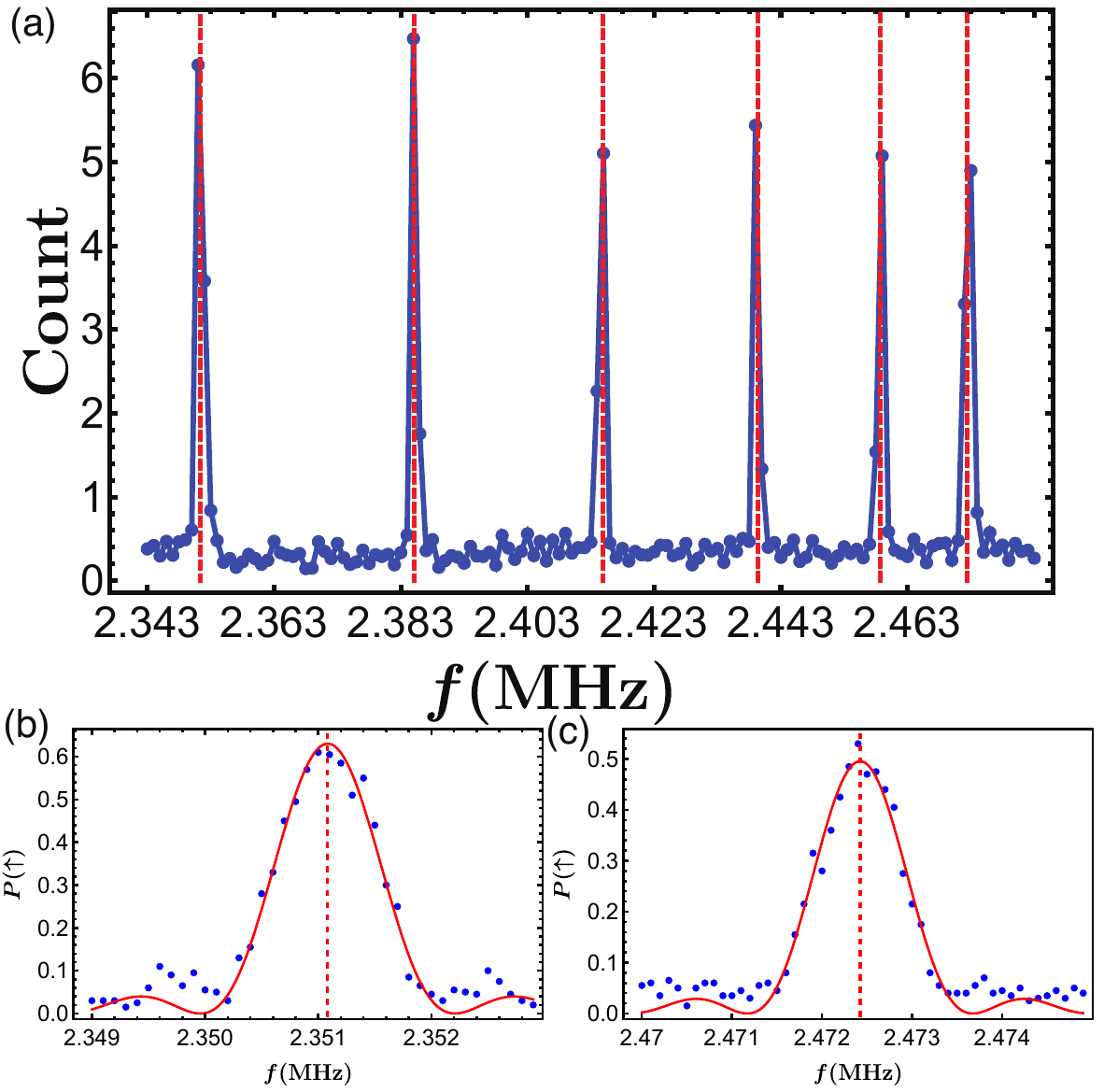}
    \caption{\label{fig:fit_spacing}(a) Measured collective mode frequencies in the $x$ direction (blue) and the theoretical values computed from the fitted ion spacings (red dashed lines). This gives us a fitting result of $\Delta z=(5.910,\,5.142,\,4.983,\,5.142,\,5.910)\,\mu$m for $N=6$ ions. (b) Fine spectrum of the lowest mode. (c) Fine spectrum of the highest mode.}
\end{figure}

In the above section, we see that the local phonon frequencies and the phonon hopping rates are determined by the inter-ion spacings as well as the transverse trap frequency. While the trap frequency can be measured accurately, the ion spacings are read out from the CCD camera and thus have larger errors limited by its resolution of about $0.3\,\mu\mathrm{m}/\mathrm{pixel}$. Note that for $N$ ions in a linear configuration, we have $N-1$ independent spacings, and that they fully determine the $N-1$ collective mode frequencies by simply diagonalizing the motional Hamiltonian Eq.~(\ref{eq:motional_RWA}) (the center-of-mass mode is just $\omega_x$) which can be measured at high accuracy. Therefore we add a step to fit the inter-ion spacings using these collective mode frequencies, as shown in Fig.~\ref{fig:fit_spacing}(a). In Fig.~\ref{fig:fit_spacing}(b) and Fig.~\ref{fig:fit_spacing}(c) we perform a finer scan for the lowest and the highest collective phonon modes which are used at higher accuracy for setting experimental parameters.

\subsection{Rabi frequencies}
\begin{figure}
    \centering
    \includegraphics[width=\textwidth]{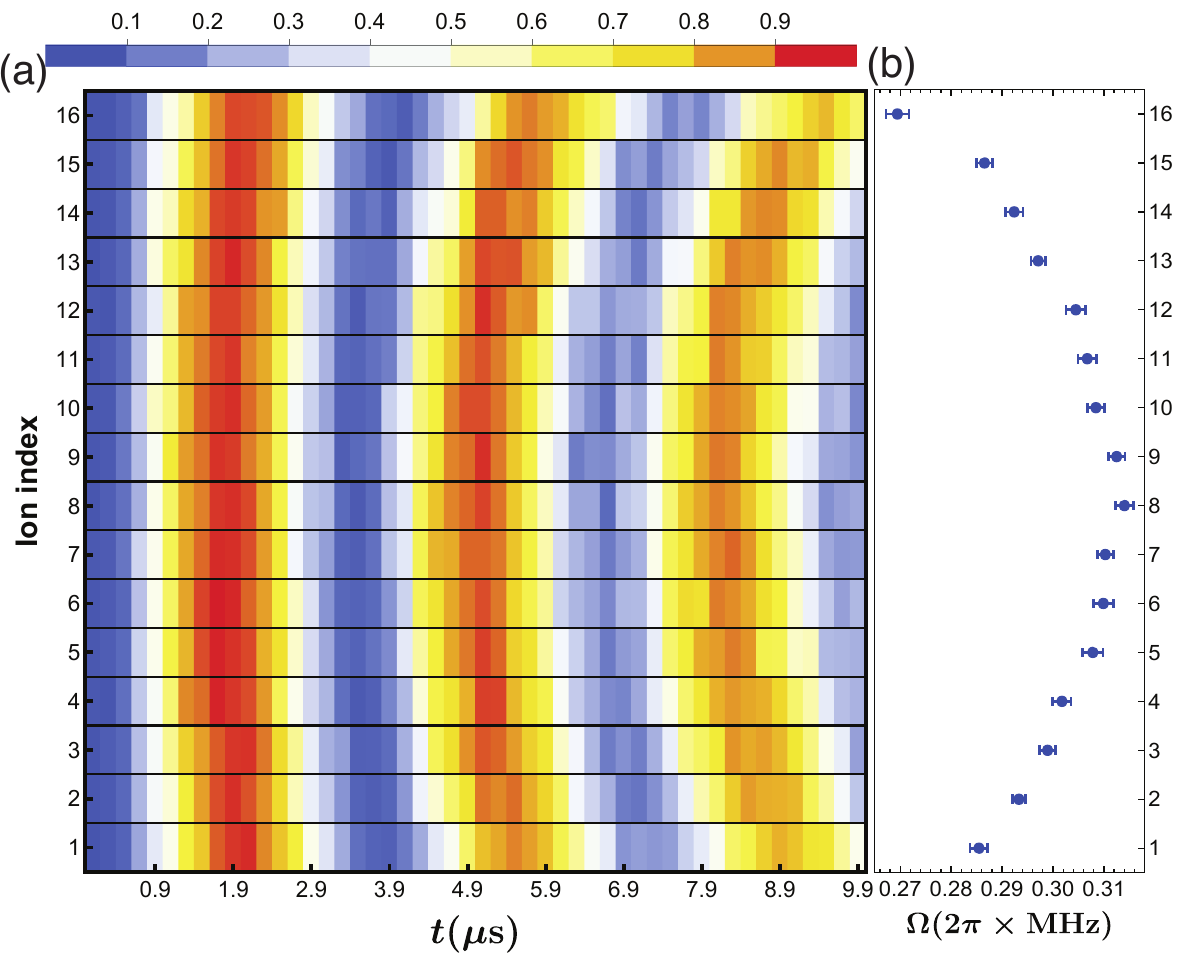}
    \caption{\label{fig:Rabi} (a) Carrier Raman Rabi oscillation for a 16-ion chain. The color map along each horizontal line corresponds to the Rabi curve for an ion with different colors representing the population of the bright state. (b) The fitted Rabi rates for individual ions. Error bars represent one standard deviation.}
\end{figure}

We calibrate the carrier and the sideband Rabi frequency by driving the Rabi oscillation for the carrier and the blue/red sidebands of a single trapped ion. However, when applying the same laser driving to a long chain, there will be nonuniformity due to the finite Gaussian profile of the laser beams. This can be seen by driving the carrier Rabi oscillation of all the ions, as shown in Fig.~\ref{fig:Rabi}. The ratio of the Rabi rates between the 16th and the 8th ion is about $86\%$. When simulating the RH model Hamiltonian, in principle we can include this nonuniformity into the theoretical model for the equilibrium and dynamical properties to compare with the experiment, but it seems not to be a dominant error source as described in the following sections, so we do not perform this correction in this work. As for the Raman $\pi$ or $\pi/2$ pulses, such a $14\%$ error in the Rabi rate leads to about $2\%$ SPAM error. If higher fidelity is needed, composite pulses can be used to suppress this amplitude error \cite{brown2004arbitrarily}.

\subsection{AC Stark shift}
\begin{figure}
    \centering
    \includegraphics[width=0.8\textwidth]{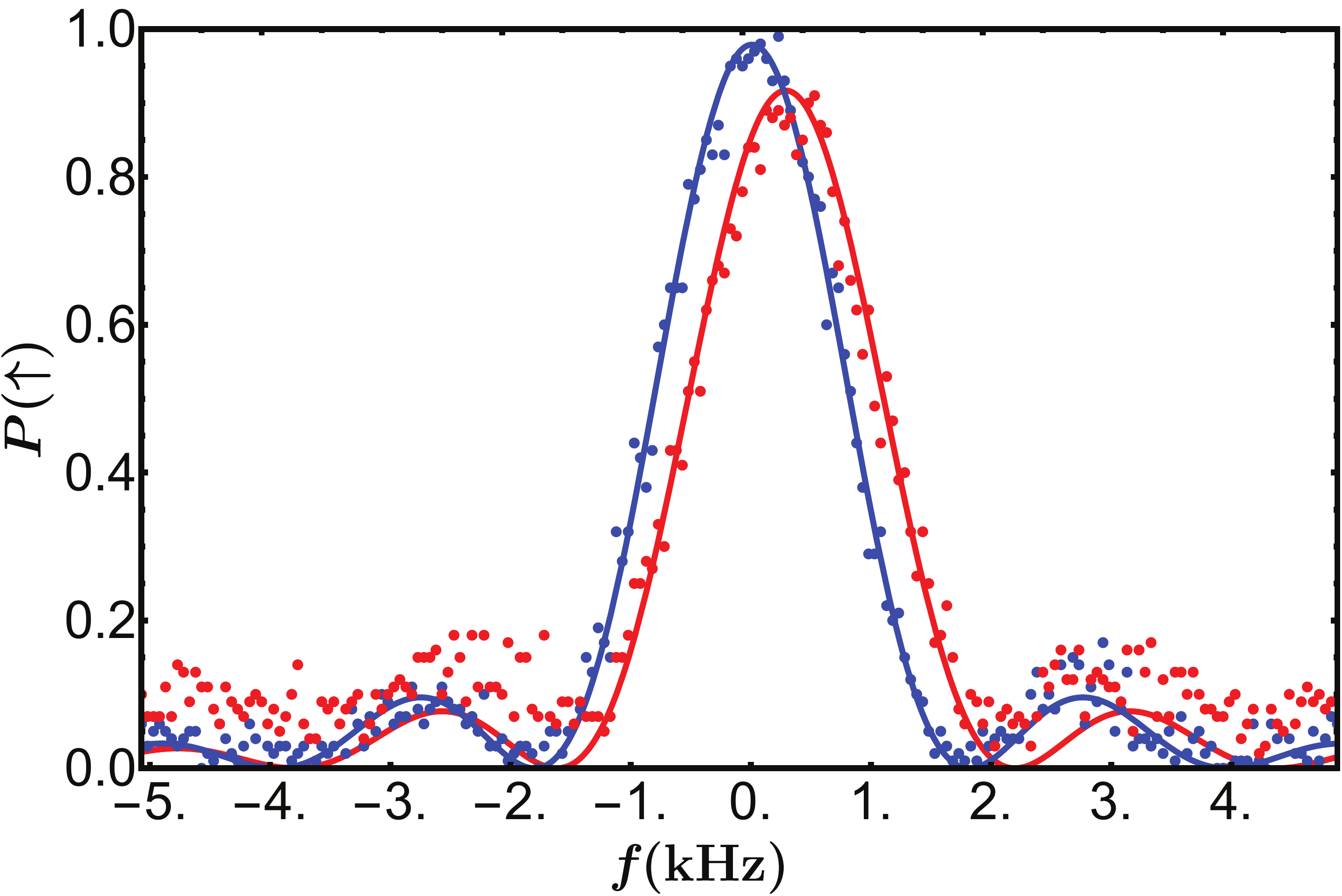}
    \caption{\label{fig:ACStark}Spectrum of a microwave near the carrier qubit transition when the red and blue sideband Raman lasers are turned off (blue) or on (red). The difference in their peaks gives the fitted AC Stark shift of about $270\,$Hz. Here the intensity of the Raman beams is set to the largest available value and is higher than what we use in the main text, so the involved AC Stark shift will be smaller.}
\end{figure}

The AC Stark shift from the Raman beams will change as we tune the laser intensity. Therefore in principle we should compensate this shift by adjusting the laser detuning accordingly during the quench process. Fortunately, as mentioned above, our $355\,$nm pulsed laser has a specially design repetition rate such that the AC Stark shift due to the blue and the red sidebands can nearly cancel each other. Here we measure the AC Stark shift by scanning the spectrum of a microwave drive on the carrier qubit transition. As shown in Fig.~\ref{fig:ACStark}, the total AC Stark shift involved in this experiment is below $300\,$Hz, and is smaller than other experimental noise such as the slow drifts in the trap frequency.

\section{Measurement of magnetization and spin-spin correlation}
We use an EMCCD to measure $\sigma_z^i$'s for all the ions in a single shot through the resonant fluorescence under the $369.5\,$nm cyclic transition with spatial resolution. Specifically, we select $5\times 5$ pixel arrays around the precalibrated ion centers and count the collected photon numbers in these regions as the data for individual ions. We calibrate the photon counts for ions in the bright and the dark states respectively, and then choose a threshold $n_{\mathrm{th}}=70$ photons for the $500\,\mu$s detection time to optimize the detection fidelity to be about $97\%$-$98\%$.

To measure the $\sigma_x$-$\sigma_x$ correlation, we apply a global $\pi/2$ pulse around the $y$ axis to rotate the $\sigma_x$ basis into the $\sigma_z$ basis. Note that the RH Hamiltonian is realized in an interaction picture which is rotating with respect to the lab frame, hence the phase of the $\pi/2$ pulse being applied should also be changing with time. In principle, we can compute this phase from the pulse sequence and apply the desired rotation. However, as the quench time is varied up to $1\,$ms, even small errors in the AC Stark shift or trap frequency can accumulate into noticeable errors in this phase. Therefore, rather than using a precomputed phase, we scan the phase $\phi$ of the $\pi/2$ pulse to extract the spin correlation (actually its absolute value). Since the RH Hamiltonian only contains $\sigma_x$ and $\sigma_z$ terms, we expect there to be no coherent $\sigma_y$ parts in the ground state. Then the $\sigma_x$-$\sigma_x$ correlation as a function of $\phi$ can be given by $C_{ij}(\phi)=C_{ij}^{(0)}\cos^2(\phi+\phi_0)$ where $\phi_0$ represents the aforementioned phase from the interaction picture, and the amplitude $C_{ij}^{(0)}$ gives the desired correlation.

In practice, we find there to be a constant shift in the sinusoidal oscillation, so we fit
\begin{equation}\label{eq:fitting}
C_{ij}^{\mathrm{exp}} = C_{ij}^{(0)}\cos^2(\phi+\phi_0) + C,
\end{equation}
where $C$ is typically smaller than 0.05 in our experiment.
We attribute this constant shift to the detection error from the CCD camera.
Specifically, due to the spreading of the photon counts from an ion on the image, a bright ion can cause a neighboring dark ion to be detected as bright as well. Therefore the $|\downarrow\uparrow\rangle$ or $|\uparrow\downarrow\rangle$ may be detected as $|\uparrow\uparrow\rangle$ with a correlated error of about $\epsilon_c=5\%$ for the nearest neighbors, and smaller for distant pairs. Suppose an ideal detection will have probability $p_{+1}/2$ for $|\uparrow\uparrow\rangle$ and $|\downarrow\downarrow\rangle$ (note that due to the $Z_2$ symmetry, ideally these two outcomes have the same probability) and probability $p_{-1}/2$ for $|\uparrow\downarrow\rangle$ and $|\downarrow\uparrow\rangle$, with $p_{+1}+p_{-1}=1$. Ideally the measured spin-spin correlation $\langle\sigma_x^i\sigma_x^j\rangle - \langle\sigma_x^i\rangle \langle\sigma_x^j\rangle$ will be $p_{+1}-p_{-1}$, while after taking into account the crosstalk error, the correlation, up to the linear order in $\epsilon_c$, becomes $p_{+1}-p_{-1}(1-2\epsilon_c)=(1-\epsilon_c)(p_{+1}-p_{-1})+\epsilon_c$. In other words, we have a reduced oscillation amplitude by $(1-\epsilon_c)$ together with a constant shift of $\epsilon_c$. Also an independent error $\epsilon_0$ for individual qubit detections will influence the measured correlation. Following the similar derivation, we see that now the correlation becomes $(1-4\epsilon_0)(p_{+1}-p_{-1})$ up to the linear order in $\epsilon_0$, hence a reduced amplitude by $(1-4\epsilon_0)$ without a constant shift.

To sum up, by extracting the spin-spin correlation using the oscillation amplitude with respect to $\phi$, we can eliminate the influence of a constant shift, which is particularly useful for the low-$g$ regime where the correlation by itself is low. The measured correlation is reduced by a small fraction due to the detection errors, which in principle can be corrected using the known error rates. However, since such a global factor does not affect our observation of the quantum phase transition qualitatively, in this experiment we do not perform this correction.

\section{Quantum phase transition}
\subsection{Mean-field analysis}
Let us consider the mean-field solution of the RH Hamiltonian [Eq.~(1) of the main text]. Without the spin-phonon coupling term ($g=0$), we can diagonalize the phonon Hamiltonian as the collective modes $b_k$ ($k=0,\,1,\,\cdots,\,N-1$) with frequencies $\delta_k$ and mode vectors $v_{ik}$ ($i=1,\,2,\,\cdots,\,N$). Then at finite $g$ we have
\begin{equation}
H = \sum_i \frac{\omega_0}{2} \sigma_z^i + \sum_k \delta_k b_k^\dag b_k + \sum_i g \sigma_x^i \sum_k v_{ik} (b_k + b_k^\dag).
\end{equation}
In this experiment, in order to observe strong transition signal, we set the lowest phonon mode $\delta_0$ to be small (note that for a stable ground state to exist, the frequencies of all the phonon modes must be positive, which is an additional requirement for studying the equilibrium phase transition but not for general dynamics of the RH model), so that we expect the phase transition property to be dominated by this mode. Then effectively we have
\begin{equation}
H = \sum_i \frac{\omega_0}{2} \sigma_z^i + \delta_0 b_0^\dag b_0 + \sum_i v_{i0} g \sigma_x^i (b_0 + b_0^\dag)
\end{equation}
Now we make the mean-field approximation by assuming small fluctuation in $\sigma_x^i$ and $b_0$ using $AB \approx \langle A\rangle B + A \langle B\rangle - \langle A\rangle \langle B\rangle$. Then we get
\begin{equation}
H = \sum_i \frac{\omega_0}{2} \sigma_z^i + \sum_i v_{i0} g \sigma_x^i (\langle b_0\rangle + \langle b_0^\dag\rangle) + \delta_0 b_0^\dag b_0 + \sum_i v_{i0} g \langle\sigma_x^i\rangle (b_0 + b_0^\dag),
\end{equation}
where the irrelevant c-number part has been discarded.

In this way, we have separated the spin and the phonon parts in the Hamiltonian. Then the ground state can be solved for the phonon mode and the individual spins as single-particle problems. Specifically, the phonon ground state is a coherent state $|-\sum_i g v_{i0} \langle\sigma_x^i\rangle / \delta_0 \rangle$ which gives $\langle b_0\rangle = \langle b_0^\dag\rangle = -\sum_i g v_{i0} \langle\sigma_x^i\rangle / \delta_0$, and the $i$-th spin ground state is at an angle of $\theta_i=\tan^{-1} (\omega_0 / 4 v_{i0} g \langle b_0 \rangle)$ with respect to the $x$ axis, which gives $\langle \sigma_x^i \rangle = -\cos \theta_i$. Combining these two results, we get
\begin{equation}
\langle b_0\rangle = \sum_i \frac{4 g^2 v_{i0}^2}{\delta_0} \frac{\langle b_0\rangle}{\sqrt{\omega_0^2 + 16 g^2 v_{i0}^2 \langle b_0\rangle^2}}.
\end{equation}

A trivial solution $\langle b_0\rangle = 0$ always exists. For there to be a nonzero $\langle b_0\rangle$, we need
\begin{equation}
1 = \sum_i \frac{4 g^2 v_{i0}^2}{\delta_0} \frac{1}{\sqrt{\omega_0^2 + 16 g^2 v_{i0}^2 \langle b_0\rangle^2}}.
\end{equation}
Note that the RHS is decreasing as $|\langle b_0 \rangle|$ increases. Therefore for a nontrivial solution to exist, we want $\mathrm{RHS}> 1$ when $\langle b_0 \rangle=0$, that is, $1 <\sum_i (4 g^2 v_{i0}^2/\delta_0\omega_0) = 4 g^2/\delta_0\omega_0$. Thus we obtain the critical coupling strength $g_c^\mathrm{mf} = \sqrt{\omega_0 \delta_0} / 2$.

Next we consider the influence of the other phonon modes at higher frequencies. Once the spins get nonzero $\sigma_x^i$ components, they will drive the other phonon modes and hence these modes will also acquire nonzero amplitudes which will in turn influence the spin states. Under the mean-field approximation, we can similarly write
\begin{equation}
\langle b_k\rangle = \sum_i \frac{4 g^2 v_{ik}}{\delta_k} \frac{\sum_l v_{il} \langle b_l\rangle}{\sqrt{\omega_0^2 + 16 g^2 (\sum_l v_{il} \langle b_l\rangle)^2}}.
\end{equation}
Let us consider the regime near the critical point $g_c^{\mathrm{mf}}$ where $\langle b_k\rangle$ are all close to zero so that they can be neglected in the denominator. Then we have
\begin{equation}
\langle b_k\rangle = \sum_i \frac{4 g^2 v_{ik}}{\delta_k \omega_0} \sum_l v_{il} \langle b_l\rangle = \frac{4 g^2 }{\delta_k \omega_0} \langle b_k\rangle
\end{equation}
For $g<g_c^\mathrm{mf}\equiv\sqrt{\omega_0 \delta_0} / 2$, the only consistent solution is $\langle b_k\rangle=0$, while for $g>g_c^\mathrm{mf}$, $\langle b_0\rangle$ will increase until the term in the denominator cannot be neglected, which will thus generates nonzero $\langle b_k\rangle$ for $k\ne 0$. Therefore the meanfield phase transition point is still $g_c^\mathrm{mf}\equiv\sqrt{\omega_0 \delta_0} / 2$. Note that this meanfield solution does not necessarily reflect the exact transition point. In the next section we use DMRG to find an approximate ground state and then extract the critical point numerically.
\subsection{Numerical results from DMRG}
We can use the DMRG algorithm to approximate the properties of the ground state, assuming low entanglement in the 1D system. This approximation may not be valid near the critical point, but it shall correctly capture the change of order parameters between the two phases.

\begin{figure}
    \centering
    \includegraphics[width=0.9\textwidth]{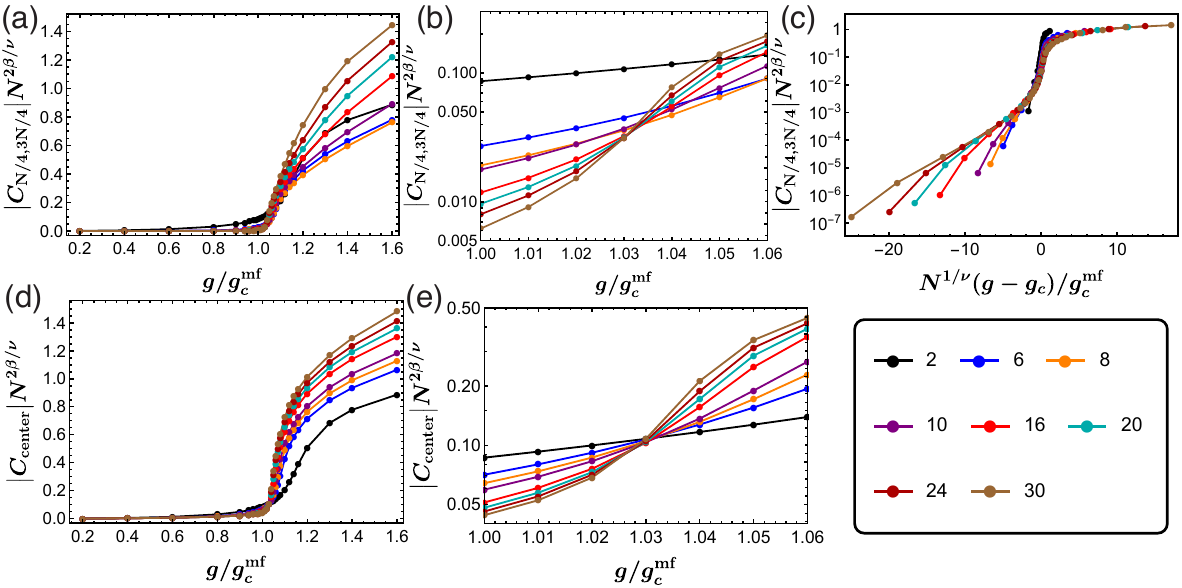}
    \caption{\label{fig:S_DMRG} Numerical results from DMRG for the quantum phase transition. (a) Spin-spin correlation over half a chain $C_{N/4,3N/4}\equiv \langle\sigma_x^{N/4}\sigma_x^{3N/4}\rangle - \langle\sigma_x^{N/4}\rangle \langle\sigma_x^{3N/4}\rangle$ vs. spin-phonon coupling $g$ for 2-30 ions with uniform spacing. We normalize $g$ by the mean-field critical point $g_c^\mathrm{mf}$ and scale the correlation by $N^{2\beta/\nu}$. (b) Semi-log plot for (a). We can extract a critical point $g_c\approx 1.03 g_c^{\mathrm{mf}}$. (c) We further present the horizontal axis as $N^{1/\nu}(g-g_c)/g_c^{\mathrm{mf}}$. Curves for different system sizes overlap near the transition point. (d,e) Similar plot as (a,b) using the nearest-neighbor correlation. We extract the same transition point as before.}
\end{figure}

We use local phonon cutoff of 10 and a bond dimension up to 30 for the numerical results presented in Fig.~\ref{fig:S_DMRG}. Because of the $Z_2$ symmetry of the Hamiltonian, we only consider the subspace with the same parity as the initial state $|\downarrow,0\rangle^{\otimes N}$. According to Ref.~\cite{flottat2016quantum}, the spin correlation over half a chain $C_{N/4,3N/4}$ can be used as an order parameter for this phase transition. Let us assume a uniform ion chain with phonon hopping rates $t_{ij}=2\pi\times 26/|i-j|^3\,$kHz, which corresponds to about $d=5.4\,\mu$m spacing and transverse trapping potential $\omega_x=2\pi\times 2.5\,$MHz. Then we can compute the local mode frequencies following Eq.~(\ref{eq:local_freq_low_order}). As described above, to get significant phase transition signal, we set the lowest collective mode frequency in the interaction picture $\delta_0$ to be small. Here we choose $\delta_0=2\pi\times 2\,$kHz. Also, Ref.~\cite{flottat2016quantum} suggests that the difference between the RH model and the JCH model is most prominent when the spin frequency and the local phonon frequencies in the interaction picture [Eq.~(1) of the main text] are equal, hence we set the spin frequency $\omega_0$ to be equal to the local mode frequency $\omega_{N/2}$ for the central ion. In Fig.~\ref{fig:S_DMRG}(a) we plot the order parameter vs. spin-phonon coupling $g$ for various system sizes up to 30 ions. We normalize the coupling $g$ by the mean-field critical point $g_c^\mathrm{mf}$ to compensate the finite size effect, and we scale the spin-spin correlation by $N^{2\beta/\nu}$ following Ref.~\cite{flottat2016quantum} where $\beta=1/8$ and $\nu=1$ are two critical exponents. As we can see, the transition signal becomes increasingly sharper as the ion number $N$ increases, and from the semi-log plot in Fig.~\ref{fig:S_DMRG}(b) we clearly see that these curves (apart from small system sizes due to the edge effect) intersect at the transition point $g_c\approx 1.03 g_c^{\mathrm{mf}}$. Furthermore, in Fig.~\ref{fig:S_DMRG}(c) we present the horizontal axis as $(g-g_c)N^{1/\nu}$ and observe that the curves for different system sizes overlap with each other near the critical point, which indicates the correct critical exponents. Finally, we note that the correlation between the two central ions can also be used to extract the phase transition signal. As shown in Fig.~\ref{fig:S_DMRG}(d) and (e), the rescaled nearest-neighbor spin-spin correlation again intersect at the critical point $g_c\approx 1.03 g_c^{\mathrm{mf}}$ for different system sizes (this time with smaller edge effects), similar to the case of the long-range correlation.

\subsection{Experimental parameters}
Here we specify the detailed parameters used in Fig.~2 of the main text. As mentioned above, we calibrate the ion spacings from the measured frequencies of the collective phonon modes, and then set the lowest collective mode in the interaction picture to be $\delta_0=2\pi\times 2\,$kHz and the spin frequency $\omega_0$ to be equal to the local mode frequency of the central ion.

For $N=2$ ions, we measure the collective phonon mode frequencies as $\omega_k=2\pi\times (2.3995,\,2.4577)\,$MHz and fit the ion spacing $\Delta z = 5.262 \,\mu$m. Then we set $\delta_b=2\pi\times  88\,$kHz and $\delta_r=-2\pi\times 32.5\,$kHz as the detuning of the blue and the red sidebands.

For $N=6$ ions, we measure the collective phonon mode frequencies as $\omega_k=2\pi\times (2.3527,\,2.386,\,2.415,\,2.439,\,2.459,\\2.4732)\,$MHz and fit the ion spacing $\Delta z = (5.847,\,5.164,\,4.990,\,5.164,\,5.847)\,\mu$m. Then we set $\delta_b=2\pi\times 171.49\,$kHz and $\delta_r=-2\pi\times73.54 \,$kHz as the detuning of the blue and the red sidebands.

For $N=10$ ions, we measure the collective phonon mode frequencies as $\omega_k=2\pi\times (2.3590,\,2.378,\,2.393,\,2.409,\,2.423,\\2.435,\,2.446,\,2.455,\,2.462,\,2.4675)\,$MHz and fit the ion spacing $\Delta z = (7.188,\,6.071,\,5.596,\,5.427,\,5.250,\,5.427,\,5.596,\\6.071,\,7.188) \,\mu$m. Then we set $\delta_b=2\pi\times 152.96\,$kHz and $\delta_r=-2\pi\times 67.94\,$kHz as the detuning of the blue and the red sidebands.

For $N=14$ ions, we measure the collective phonon mode frequencies as $\omega_k=2\pi\times (2.3582,\,2.373,\,2.387,\,2.400,\,2.412,\\2.424,\,2.434,\,2.444,\,2.453,\,2.461,\,2.468,\,2.473,\,2.478,\,2.4821)\,$MHz and fit the ion spacing $\Delta z = (7.585,\,6.385,\,5.807,\\5.477,\,5.267,\,5.168,\,5.119,\,5.168,\,5.267,\,5.477,\,5.807,\,6.385,\,7.585)\,\mu$m. Then we set $\delta_b=2\pi\times 179.36\,$kHz and $\delta_r=-2\pi\times 72.40\,$kHz as the detuning of the blue and the red sidebands.

For $N=16$ ions, we measure the collective phonon mode frequencies as $\omega_k=2\pi\times (2.3524,\,2.365,\,2.377,\,2.388,\,2.399,\\2.410,\,2.419,\,2.428,\,2.437,\,2.444,\,2.451,\,2.457,\,2.463,\,2.468,\,2.471,\,2.4744)\,$MHz and fit the ion spacing $\Delta z = (7.764,\\6.737,\,6.018,\,5.644,\,5.437,\,5.283,\,5.195,\,5.183,\,5.195,\,5.283,\,5.437,\,5.644,\,6.018,\,6.737,\,7.764)\,\mu$m. Then we set $\delta_b=2\pi\times 175.22\,$kHz and $\delta_r=-2\pi\times 72.90\,$kHz as the detuning of the blue and the red sidebands.

\subsection{Experimental results}
\begin{figure}
    \centering
    \includegraphics[width=0.9\textwidth]{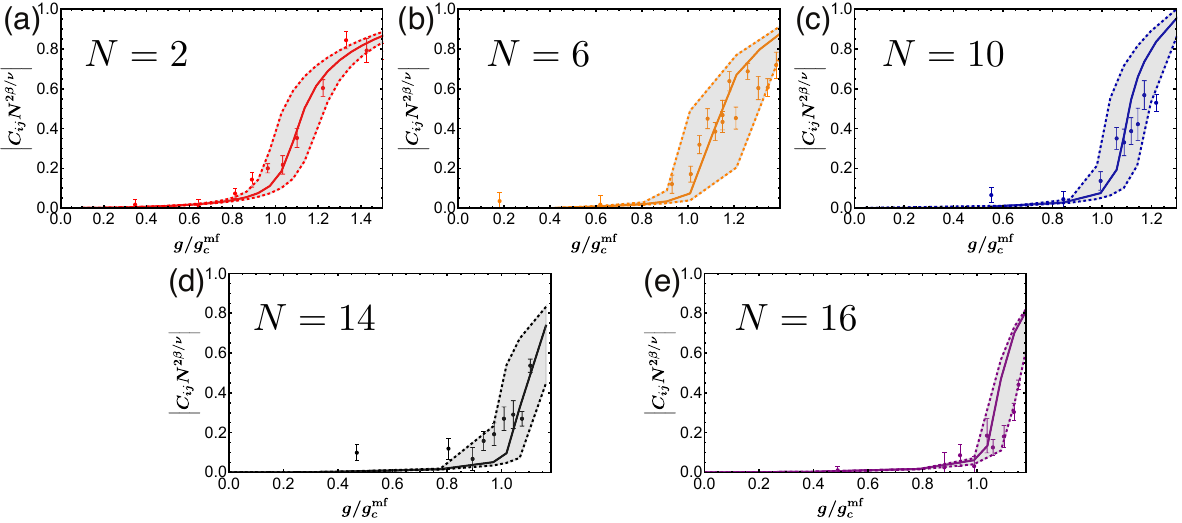}
    \caption{\label{fig:S_2F} Individual plots for the experimental data and the corresponding theoretical curves in Fig.~2(f) of the main text. The shaded region between the dashed curves represents a shift of $\pm300\,$Hz in the trap frequency.}
\end{figure}

In Fig.~\ref{fig:S_2F} we plot the experimental data in Fig.~2(f) of the main text for each ion number $N$ individually, together with their corresponding theoretical values. From these plots we can see more clearly that the theoretical and the experimental results agree with each other for different system sizes.

\subsection{Adiabaticity in slow quench across the critical point}
\begin{figure}
    \centering
    \includegraphics[width=0.9\textwidth]{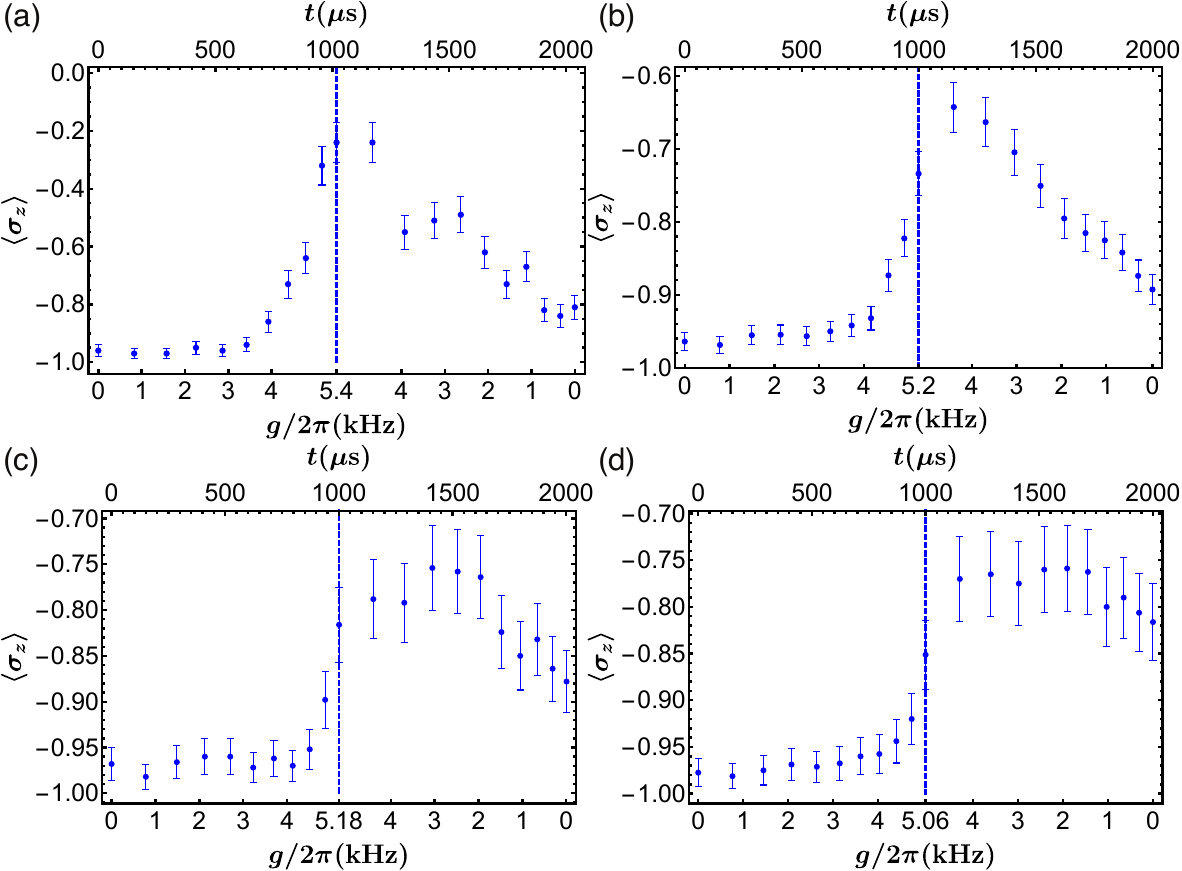}
    \caption{\label{fig:adiabatic} We follow the sequence in the main text to slowly tune the spin-phonon coupling to the highest value, and then reverse the sequence to tune it back to zero. The four subplots are the average magnetization $\langle\overline{\sigma_z}\rangle$ for 2, 6, 10, 16 ions, respectively. Initially the ions are prepared in $|\downarrow\rangle^{\otimes N}$ with $\langle\overline{\sigma_z}\rangle=-1$. As the system approaches the critical point, the magnetization increases as the orientations of the spins move away from the $z$ axis to acquire $\sigma_x$-$\sigma_x$ correlations. If the system stays in the instantaneous ground state, the average magnetization will finally go back to $-1$. Deviation can be caused by nonadiabatic excitations to higher levels and decoherence during the slow quench.}
\end{figure}

In Fig.~\ref{fig:adiabatic} we examine how well the system stays in the instantaneous ground state of the RH model during the slow quench. We follow the experimental sequence to slowly turn up the spin-phonon coupling across the critical point, and then we reverse the sequence to turn the coupling back to zero. If the system stays in the instantaneous ground state, the average magnetization $\langle\overline{\sigma_z}\rangle$ over the chain will first increase and then decrease in a symmetric way, going back to its initial value of $-1$. As we can see in Fig.~\ref{fig:adiabatic}, for small ion numbers, the adiabatic condition is fulfilled relatively well and the final average magnetization can be close to its initial value. As the ion number increases, the deviation from the instantaneous ground state becomes larger and the curve for the reverse quench is lifted higher and higher, which indicates the breakdown of the adiabatic condition as the energy gap shrinks with the increasing ion number at the critical point. Also, our direct numerical simulation for the two-ion dynamics suggests a more symmetric curve than the measured one, and that the observed flattening in the reverse quench region may be explained by a motional dephasing time on the order of hundreds of microseconds. The reason why this coherence time is reduced from the single-ion measurement is still not fully understood to us. Therefore we believe that the violation of the adiabatic condition as well as the reduced motional coherence for large ion number can be the dominant error source in our experiment for quantum phase transition.

\section{Dynamical evolution}
\subsection{Experimental parameters}
Here we specify the detailed parameters used in Fig.~3 of the main text. In this part, we set the laser detuning such that the collective phonon mode frequencies have both positive and negative values in order to obtain strong phonon excitation and hence more complicated dynamics. Different from the case of ground state phase transition, here a negative phonon frequency is valid because we are studying the dynamics from a well-defined initial state.

For $N=2$ ions, we measure the collective phonon mode frequencies as $\omega_k=2\pi\times (2.3837,\,2.4422)\,$MHz and fit the ion spacing $\Delta z = 5.266\,\mu$m. Then we set $\delta_b=2\pi\times 31.25\,$kHz and $\delta_r=-2\pi\times 27.25\,$kHz as the detuning of the blue and the red sidebands. This corresponds to spin frequency $\omega_0=2\pi\times 2\,$kHz and local phonon frequencies $\omega_i=2\pi\times(0,\,0)\,$kHz in the interaction picture, or collective mode frequencies $\delta_k=2\pi\times (-29.25,\, 29.25)\,$kHz.

For $N=4$ ions, we measure the collective phonon mode frequencies as $\omega_k=2\pi\times (2.4021,\,2.4264,\,2.4460,\,2.4607)\,$MHz and fit the ion spacing $\Delta z = (6.536,\, 6.113,\,  6.536)\,\mu$m. Then we set $\delta_b=2\pi\times 31.28\,$kHz and $\delta_r=-2\pi\times 27.28\,$kHz as the detuning of the blue and the red sidebands. This corresponds to spin frequency $\omega_0=2\pi\times 2\,$kHz and local phonon frequencies $\omega_i=2\pi\times(11.5,\,-6.5,\,-6.5,\,11.5)\,$kHz in the interaction picture, or collective mode frequencies $\delta_k=2\pi\times (-29.4,\, -5.1,\, 15.2,\, 29.3)\,$kHz.

For $N=16$ ions, we measure the collective phonon mode frequencies as $\omega_k=2\pi\times (2.3442,\, 2.357,\, 2.370,\, 2.381,\, 2.393,\\ 2.403,\, 2.413,\, 2.422,\, 2.431,\, 2.439,\, 2.446,\, 2.452,\, 2.458,\, 2.463,\, 2.467,\, 2.4700)\,$MHz and fit the ion spacing $\Delta z = (7.772,\\ 6.608,\,5.993,\,5.615,\,5.378,\,5.251,\,5.116,\,5.176,\,5.116,\,5.251,\,5.378,\, 5.615,\,5.993,\,6.608,\,7.772)\,\mu$m. Then we set $\delta_b=2\pi\times 65\,$kHz and $\delta_r=-2\pi\times 61\,$kHz as the detuning of the blue and the red sidebands. This corresponds to spin frequency $\omega_0=2\pi\times 2\,$kHz and local phonon frequencies $\omega_i=2\pi\times(51.5,\, 35.9,\, 22.9,\, 11.7,\, 2.5,\, -4.1,\, -9.3, \,
-11.1,\, -11.1,\\ -9.3,\, -4.1,\, 2.5,\, 11.7,\, 22.9, \,
35.9,\, 51.5)\,$kHz in the interaction picture, or collective mode frequencies $\delta_k=2\pi\times (-62.8,\,-50.0,\, -37.0,\,-26.0,\,-14.0,\,-4.0,\,6.2,\,15.2,\,23.8,\,31.6,\,38.9,\,45.5,\,51.4,\,56.4,\, 60.4,\,63.0)\,$kHz.

\subsection{Phonon cutoff when integrating Schrodinger equation}
\begin{figure}
    \centering
    \includegraphics[width=0.9\textwidth]{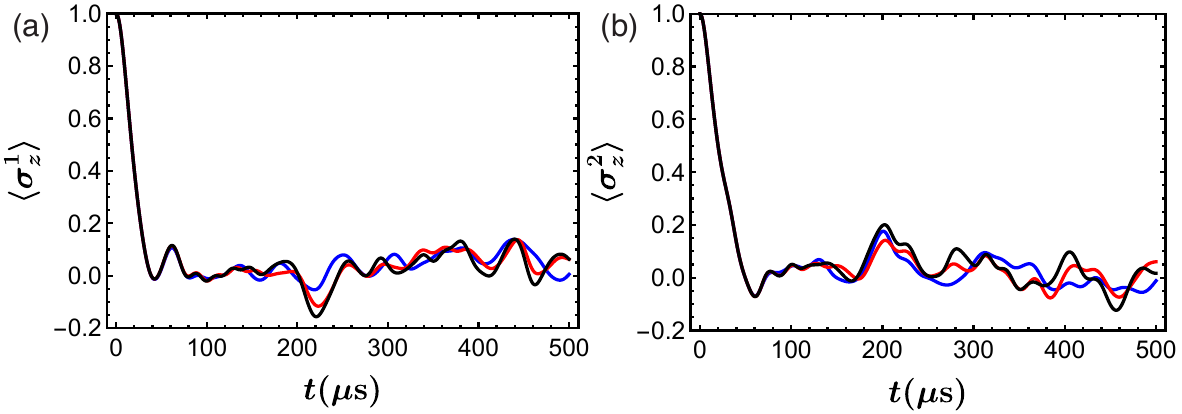}
    \caption{\label{fig:S_dynamics} Spin dynamics by integrating the Schrodinger equation for $N=4$ ions with a local phonon cutoff of $n_{\mathrm{cut}}=6$ (blue), $n_{\mathrm{cut}}=8$ (red) and $n_{\mathrm{cut}}=10$ (black). (a) The ion on the edge. (b) The ion in the center.}
\end{figure}

For small system sizes like two and four ions, we can directly integrate the Schrodinger equation to solve the dynamics of the system once we set a cutoff for the phonon modes. Suppose the phonon population follows a Poisson distribution (which is the case e.g. for a coherent state). Then for an average phonon number of about 1.5 as shown in Fig.~3 of the main text, we need a cutoff of 6 for the discarded population to be below 0.1\%. Actually, when we increase the cutoff from 6 to 8 to 10, there is still visible changes in the dynamics as shown in Fig.~\ref{fig:S_dynamics}. In the main text we use the small cutoff of $n_{\mathrm{cut}}=6$ to estimate the computational cost as $[2\times(n_{\mathrm{cut}}+1)]^N$, where $+1$ comes from the fact that the phonon number can be any integers from zero to $n_{\mathrm{cut}}$ in the truncated Hilbert space.

\begin{figure}
    \centering
    \includegraphics[width=0.5\textwidth]{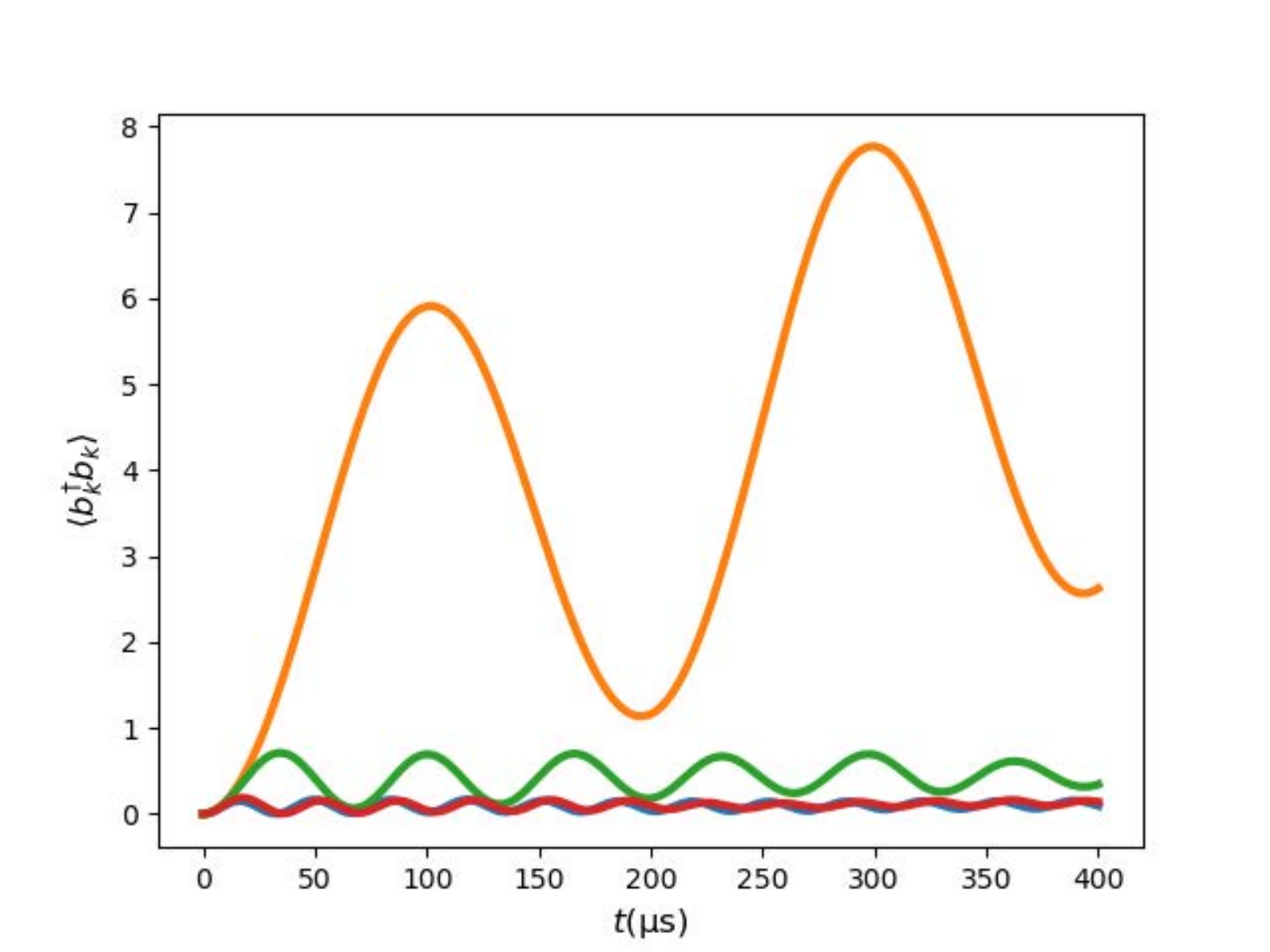}
    \caption{\label{fig:S_collective} Evolution of phonon numbers in each collective phonon mode by integrating the Schrodinger equation for $N=4$ ions with a local phonon cutoff of $n_{\mathrm{cut}}=14$. The blue, orange, green and red curves are for the collective modes in ascending order in their frequencies.}
\end{figure}

A possibly more efficient scheme is to consider the collective modes. As we show in the previous section, in the parameter regimes we are considering, the collective phonon modes have a wide distribution $\delta_k$ and some of them can be larger than the spin-phonon coupling $g$. In Fig.~\ref{fig:S_collective} we plot the evolution of the collective phonon numbers for the $N=4$ case (the blue curves in Fig.~3(g) of the main text). As we can see, the average phonon number in each collective mode does not follow $(g\delta_k)^2$ but are better estimated by $\bar{n}_k\sim(\sqrt{N}g/\delta_k)^2=(0.17,\,5.5,\,0.62,\,0.17)$. This factor of $\sqrt{N}$ may come from the fact that the collective phonon modes are coupled to all the spins with an collective enhancement effect. Now if we generalize this scaling to the $N=16$ case, we can estimate the average phonon number in each collective mode as $\bar{n}_k=(0.15,\,0.23,\,0.42,\,0.85,\,2.9,\,36,\,15,\,2.5,\,1.0,\,0.58,\,0.38,\,0.28,\,0.22,\,
0.18,\,0.16,\,0.15)$. Again if we want a truncation error below $0.1\%$ for each mode, now the cutoff should be $n_k^{\mathrm{cut}}=(2,\,3,\,3,\,5,\,9,\,56,\,28,\,9,\,5,\,4,\,3,\,3,\,3,\,2,\,2,\,2)$. This gives us an estimation for the required Hilbert space dimension as $2^N\prod_k (n_k^\mathrm{cut}+1)\approx 2^{57}$.

\subsection{Qualitative understanding by Holstein–Primakoff approximation}
In this experiment, when studying the dynamics of the RH model, we are initializing all the spins to $|\uparrow\rangle$ and all the phonon modes to $|0\rangle$. In the limit where each $\langle\sigma_z^i\rangle$ only deviates weakly from the initial value of one, we can perform the Holstein–Primakoff (HP) transformation \cite{PhysRev.58.1098} and approximate each spin by a bosonic mode
\begin{equation}
\sigma_z^i = 1 - 2 s_i^\dag s_i, \quad \sigma_+ \approx s_i, \quad \sigma_- \approx s_i^\dag.
\end{equation}

Under this approximation, the Hamiltonian can be rewritten as a quadratic form of $2N$ bosonic modes
\begin{equation}
H = \sum_i \left[\frac{\omega_0}{2} (1-2s_i^\dag s_i) + \omega_i a_i^\dag a_i + g (s_i+s_i^\dag) (a_i + a_i^\dag)\right] + \sum_{i<j} t_{ij} (a_i^\dag a_j + a_j^\dag a_i),
\end{equation}
which leads to a set of Heisenberg equations
\begin{align}
\frac{d s_i}{dt} =& i \left[\omega_0 s_i - g(a_i+a_i^\dag)\right], \\
\frac{d s_i^\dag}{dt} =& i \left[-\omega_0 s_i^\dag + g(a_i+a_i^\dag)\right], \\
\frac{d a_i}{dt} =& i \left[-\omega_i a_i - g(s_i+s_i^\dag) - \sum_{j\ne i} t_{ij} a_j\right], \\
\frac{d a_i^\dag}{dt} =& i \left[\omega_i a_i^\dag + g(s_i+s_i^\dag) + \sum_{j\ne i} t_{ij} a_j^\dag\right].
\end{align}

We can further express this set of linear equations in a matrix form: $d\vec{v}/dt = A\vec{v}$ where $\vec{v}\equiv (\vec{u}_1,\cdots,\vec{u}_N)^T$ and $\vec{u}_i=(s_i,s_i^\dag,a_i,a_i^\dag)$ ($i=1,\cdots,N$). The time evolution of each operator can thus be fully determined (assuming weak excitation $\langle s_i^\dag s_i\rangle(t) \ll 1$) by the exponentiation of the numerical matrix $A$. Specifically, let us denote $B(t)=e^{At}$. Then for the observables we are considering, $\langle\sigma_z^i\rangle(t)=1-2\langle s_i^\dag s_i\rangle(t)=1-2 \sum_j [B_{s_i^\dag,s_j}(t) B_{s_i,s_j^\dag}(t) + B_{s_i^\dag,a_j}(t) B_{s_i,a_j^\dag}(t)]$ where $B_{s_i,s_j^\dag}(t)$ is the matrix element of $B(t)$ at the row for $s_i$ and the column for $s_j^\dag$ and similarly for the other expressions. In this derivation we use the fact that the initial state is vacuum for all the modes $s_i$ and $a_i$.

\begin{figure}
    \centering
    \includegraphics[width=0.9\textwidth]{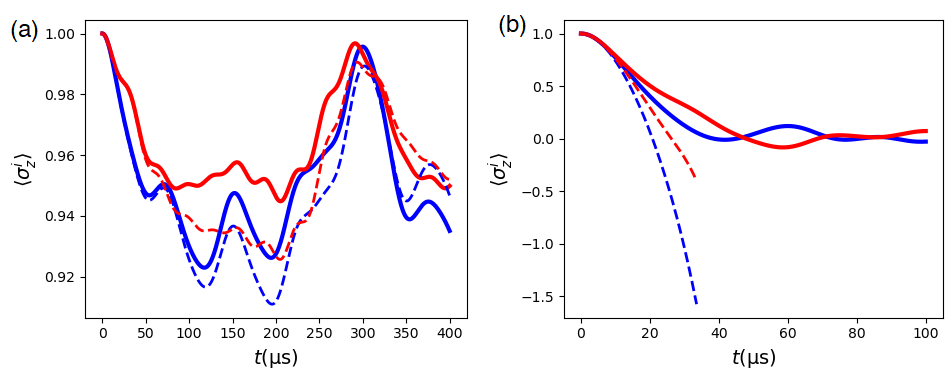}
    \caption{\label{fig:S_HP} Spin dynamics for the edge (blue) and the central (red) ions at $N=4$. Solid curves are exact solutions by integrating the Schrodinger equation, while the dashed curves are from the HP approximation. (a) Weak spin-phonon coupling $g=2\pi\times 1\,$kHz (red curves in Fig.~2(d) and 2(e) of the main text). (b) Strong spin-phonon coupling $g=2\pi\times 6\,$kHz (blue curves in Fig.~2(d) and 2(e) of the main text).}
\end{figure}

As an example, we show the comparison between the exact solution for $N=4$ by numerically integrating the Schrodinger equation and the result of HP approximation in Fig.~\ref{fig:S_HP}. As we can see, at low spin-phonon coupling, the approximation works reasonably well with the correct qualitative behavior together with many quantitative features. In this case, the deviation of the spin state from $\langle\sigma_z^i\rangle=1$ is small so that the approximation condition holds up to long evolution time. On the other hand, at strong spin-phonon coupling, the discrepancy between the two methods quickly increases as the HP approximation condition breaks down. Qualitatively, these two distinct behaviors can be explained by the stability of the coupled bosonic system under the HP approximation. Specifically, we find that in the weak coupling regime, all the eigenvalues of the matrix $A$ are purely imaginary, thus only weak oscillation around the initial state is observed; on the other hand, in the strong coupling regime, some of the eigenvalues now take real positive values, which leads to an exponential increase in the excitation number and finally the break down of the HP approximation.

Similarly, by evaluating the matrix $A$ for the $N=16$ case using the parameters listed above, we find that $g=2\pi\times 1\,$kHz and $g=2\pi\times 6\,$kHz locate in these two different regimes, which explains their different qualitative behaviors. Note that this does not contradict with our claim that the experimentally simulated dynamics is challenging for classical computers because we have reached the strong coupling regime where such approximations are not valid for long-time evolution.

% \bibliographystyle{apsrev4-1-title}
% \bibliography{reference}
